\documentstyle[12pt]{article}
\title{
{\vspace{-3cm} \normalsize \hfill MS-TPI-92-27
                                            }\\[25mm]
Investigation of the biconjugate gradient algorithm for the inversion of
fermion matrices}

\author{Markus Plagge \\
        Institut f\"ur Theoretische Physik I,
        Universit\"at M\"unster\\
        Wilhelm-Klemm-Str.~9, D-4400 M\"unster, Germany}
\date{December 7, 1992}

\newcommand{\be}{\begin{equation}}
\newcommand{\ee}{\end{equation}}

\begin{document}
\maketitle

\begin{abstract} \normalsize
An algorithm for the numerical inversion of large matrices, the
biconjugate gradient algorithm (BGA), is investigated in view of its
use for Monte Carlo simulations of fermionic field theories.
It is compared with the usual conjugate gradient algorithm (CGA) and the
minimal residue algorithm (MRA) within a Higgs-Yukawa model including
mirror fermions.
For this model it can be shown that BGA represents an improvement
under certain circumstances where the other two slow down.
\end{abstract}

\section{Introduction}
The most time consuming part in numerical investigations of field
theories including fermions on the lattice is the inversion of the
fermion matrix during the Monte Carlo updating.
If we call this matrix Q, then, using the Hybrid Monte Carlo (HMC)
algorithm, for each leapfrog step the equation
\begin{equation}
\label{1}
Q^+ Q x = b
\end{equation}
has to be solved.
Usually this is done by the {\it conjugate gradient algorithm} (CGA)
\cite{hes,eis,fab}, which allows the inversion of a positive definite and
Hermitean $N \times N$ matrix within at most N iteration steps.
Actually this is true only for an analytical inversion.
Due to roundoff errors during a numerical inversion one has to set an
error bound $\delta$ , which terminates the algorithm when this bound is
met.
Nevertheless at least analytically CGA is forced to converge after a
maximum number of steps.

The {\it minimal residue algorithm} (MRA) \cite{eis,oya,ros}
is another well-known method for the inversion of large matrices.
In some sense it is the opposite of CGA.
Firstly, MRA is able to treat non-Hermitean and non-positive definite
matrices, which on the one hand allows its application to more general
problems, and on the other hand allows a more flexible dealing with the
matrix that has to be inverted (see below).
Secondly, in contrast to CGA, it is not forced to converge within a predictable
number of iterations.

Because of the large amount of computer time for such iterative
inversions a lot of effort has been spent in the acceleration of this
part of the Monte Carlo algorithms.
One important possibility to reach this aim is preconditioning
\cite{ros,hoc}.
Preconditioning means the transformation of the original matrix into
another one, which can be treated more easily by the algorithm of
inversion.
If this matrix is ``closer'' to the unit matrix than the original
matrix, its inversion will be faster.

The structure of many matrices that result from the discretisation
of differential equations are convenient for the so called ``odd-even''
or ``red-black'' preconditioning \cite{hoc,gup}.
But this only works, if in equation (\ref{1}) $Q^+$ and $Q$ are
inverted separately.
Because both of these matrices are usually neither Hermitean nor
positive definite this cannot be done by CGA, but e.g.\ by MRA.
In this sense the latter is more flexible.

The effect of combining preconditioning with MRA was tested within QCD
\cite{gup} and a Higgs-Yukawa model, respectively \cite{desy92111}.
It was found that for small fermionic
hopping parameters $K$ or large fermion masses the use of MRA results
in a considerable gain of computer time.
For larger $K$ the time requirements using MRA increase drastically, due
to the fact that its convergence properties are worse than those of CGA.
In those $K$-regions it appears that only the inversion by CGA remains
feasible and therefore the odd-even preconditioning is no longer
possible.

Here the {\it biconjugate gradient algorithm} (BGA) \cite{rec,76} comes
into play. It combines the advantages of both algorithms described above.
On the one hand BGA, which can be considered as a generalisation of
CGA, is also forced to converge like the latter.
On the other hand it allows the inversion of non-Hermitean and
non-positive definite matrices and therefore the separate treatment of
$Q^+$ and $Q$ like MRA.
Investigations of some general properties of this algorithm have been done
in numerical mathematics. But to the best of my knowledge it has so far
not been considered for the inversion of fermion matrices.
This article is devoted to the examination and comparison of the three
algorithms within the $\rm U(1)_L \otimes U(1)_R$-symmetric
and the $\rm SU(2)_L \otimes SU(2)_R$-symmetric
Higgs-Yukawa model \cite{Desy 90035,Desy 90105,Desy 91}.

Before going into more detail, I would like to mention another
possibility to accelerate the iterative matrix inversion, the so
called ``educated guess'', because it is used during all our
calculations.
It consists of guessing a first approximate solution $x_1$ of equation
(\ref{1}), or corresponding equations, taking into account available
foreknown information.
In our case we suggest that the system under investigation and therefore
the fermion matrix, which is a part of it, shows a continuous behaviour
during each HMC trajectory.
Therefore using previous solutions of (\ref{1}) within the same
trajectory, one can make an extrapolation to the solution of the actual
equation, using the ``Gottlieb-trick'' \cite{got}.
Keeping this in mind, it is plausible that the quality of the guess is
strongly influenced by the step size $\epsilon$ used during the
HMC trajectory.
If $\epsilon = 0$ the solutions of two succeeding equations within a
trajectory have to be the same.
On the other hand, if $\epsilon$ is very large, they have nearly nothing
to do with each other.

\section{The Algorithm}
Let A be a complex $N \times N$ matrix, which need not to be Hermitean or
positive definite, but invertible.
To solve the equation $Ax=b$ BGA constructs five sequences of complex
N-component vectors $r_k,\bar{r}_k,p_k,\bar{p}_k,x_k$ by the following
recurrence:
\begin{eqnarray}
\label{bi0}
\alpha_k &=& \frac{(\bar{r}_k,r_k)}{(\bar{p},Ap)}   \nonumber\\
 r_{k+1} &=& r_k - \alpha_k A p_k                   \nonumber\\
 \bar{r}_{k+1} &=& \bar{r}_k - \alpha_k^* A^+ \bar{p}_k \nonumber\\
 x_{k+1} &=& x_k + \alpha_k p_k                        \nonumber\\
 \beta_k &=& \frac{(\bar{r}_{k+1},r_{k+1})}{(\bar{r}_k,r_k)}  \nonumber\\
 p_{k+1} &=& r_{k+1} + \beta_k p_k                       \nonumber\\
 \bar{p}_{k+1} &=& \bar{r}_{k+1} + \beta_k^* \bar{p}_k \,.
\end{eqnarray}
The vectors are set initially to $r_1=b-Ax_1$, where $x_1$ is some guess
of the solution, and  $\bar{r}_1=\bar{p}_1=p_1=r_1$. The scalar product is
defined as $(a,b) = \sum_{i=1}^{N} a_i^* b_i$.

For this recurrence the {\it biorthogonality} condition,
\begin{equation}
\label{bi1}
(\bar{r}_i,r_j) = (r_i,\bar{r}_j) = 0
\hspace{1.0cm}  \forall_{j<i} \,,
\end{equation}
and the {\it biconjugacy} condition,
\begin{equation}
\label{bi2}
(\bar{p}_i,Ap_j) = (Ap_i,\bar{p}_j) = 0
\hspace{1.0cm}  \forall_{j<i} \,,
\end{equation}
can be derived in the same way as it was done for real matrices A in
\cite{76}.
The only difference is that now $\alpha_i$ and $\beta_i$ are complex
numbers and have to be treated differently for $r_i,p_i$ and
$\bar{r}_i,\bar{p}_i$, respectively.

{}From the conditions above it follows that the $r_i$ form a sequence of
linearly independent vectors, unless the algorithm breaks down, if one
of the denominators in (\ref{bi0}) vanishes.
Therefore BGA must terminate after $m \le N$ steps with $r_{m+1}=0$,
if rounding errors are neglected for the moment.
The sequence $x_i$ consequently leads to $x_{m+1}$ being the solution
of $Ax=b$.

The above mentioned break down, owing to vanishing denominators,
apparently occurs quite rarely \cite{rec} and has indeed never been
observed to occur during the tests within the Higgs-Yukawa model.

Due to roundoff errors during the numerical inversion some error bound
$\delta$ must be set as for CGA.
As these errors occur in the inversion of both $Q^+$ and $Q$, we have a
complicated error propagation.
In order to achieve the same precision as for the one-step inversion by
CGA, we decrease the error bound for the separated inversions by a
factor of 100.
We have checked this for both MRA and BGA at several points and
found that it is more than sufficient to get the desired precision.

\section{Results}
Before comparing the runtime of the three algorithms, one can make
some qualitative predictions.
In order to do this let us have a closer look at the fermion matrices of
Yukawa models with mirror fermions.
For more details concerning the lattice action etc.\ see
\cite{Desy 90035,Desy 90105,Desy 91}.
The fermion matrix can be written as:
\begin{eqnarray}
Q(\phi)_{yx} &=&
\delta_{yx} \left(\begin{array}{cccc} G_{\psi}\phi_x^+ & 0 & {\bf 1} & 0 \\
                                      0 & G_{\psi} \phi_x & 0 & {\bf 1}  \\
                                      {\bf 1} & 0 & G_{\chi}\phi_x & 0   \\
                                      0 & {\bf 1} & 0 & G_{\chi}\phi_x^+
                    \end{array} \right) \nonumber\\
             & &  - K \sum_{\mu}
 \delta_{y,x+\mu} \left( \begin{array}{cccc} 0 & \Sigma_{\mu} & {\bf 1} & 0  \\
                                        \bar{\Sigma}_{\mu} & 0 & 0 & {\bf 1} \\
                                        {\bf 1} & 0 & 0 & \Sigma_{\mu}       \\
                                        0 & {\bf 1} & \bar{\Sigma}_{\mu} & 0
                  \end{array} \right) \,,
\end{eqnarray}
where $G_\psi,G_\chi$ are Yukawa couplings, $\phi$ is the Higgs field,
$K$ the fermion hopping parameter, already mentioned above, and ${\bf 1}$
the two-dimensional unit matrix.
The sum over $\mu$ goes over all eight lattice directions.
The $\Sigma_\mu$ are connected to the Pauli matrices by $\Sigma_\mu =
-\bar{\Sigma}_\mu= -i \sigma_\mu$ for
$\mu=1,2,3$ and $\Sigma_4=\bar{\Sigma}_4 ={\bf 1}$.
For negative indices the definition is
$\Sigma_\mu \equiv -\Sigma_{-\mu}$.

We distinguish three different kinds of parameters, which will be
considered separately.
The Yukawa couplings and the fermionic hopping parameter $K$ are called
{\it direct} parameters, because they enter $Q$ explicitly.
Parameters like the scalar hopping parameter $\kappa$ and the quartic
scalar self-coupling $\lambda$ affect $Q$ only through the Higgs field
configuration and are therefore called {\it indirect} parameters.
Their influence on the performance of the inversion cannot be examined
as easy as that of the first one.
A third kind of parameters is called {\it algorithmic}.
Among them are the lattice size, the order of preconditioning $P$
\cite{gup,siam1}, the leapfrog step size $\epsilon$, and the error bound
$\delta$.
Overrelaxation, which introduces a further algorithmic parameter
$\omega$, is not possible here, because the relations
(\ref{bi1},\ref{bi2}) are valid only for the trivial case $\omega=1$.

\subsection{Algorithmic parameters}
The influence of the algorithmic parameters will not be examined in
great detail here.
But I would like to make some general remarks to give a qualitative
picture.
\begin{enumerate}
\item Going to the next higher order of $P$ in BGA, the amount
of time for one iteration of the matrix inversion doubles, like it
does in the MRA.
In order to get a gain the number of iterations for one inversion must
therefore decrease by a factor of more than 2.
Altough this is the case for MRA in certain regions of the
parameter space, it is usually not true for BGA.
In regions where the BGA is competitive with the other algorithms,
a factor of 1.7 or less is found.
Only in those regions where MRA still represents the fastest method
of inversion we can get an improvement by choosing some higher order of
$P$.
Because these parameter regions are not interesting for BGA we consider
only first order preconditioning in all following calculations.
\item In our calculations we have to fix the step size $\epsilon$ at
a value, where the acceptance rate for the HMC-updating is about 75\%.
Despite this one can vary $\epsilon$ in order to see how the different
algorithms behave.
As mentioned above, $\epsilon$ influences the educated guess based on the
Gottlieb-trick, and
therefore the number of iterations per inversion.
Thus it is even possible that for certain large values of $\epsilon$,
corresponding to acceptance rates far below 75\%, it will be better to
do the inversion without educated guess, simply setting $x_1$ to
zero.
During all tests it turned out that CGA is more sensitive to changes
in the quality of the educated guess, whereas BGA and MRA appear to be
more stable.
This means that it can depend on the value of $\epsilon$, which of the
algorithms shows the best performance.
If the acceptance rate is fixed to 75\% it can not in general be
foreseen whether the educated guess works better for CGA or BGA and
MRA, respectively.
\item Concerning error bound $\delta$ and lattice size BGA and
MRA behave quite similarly.
The larger the lattice or the smaller the error bound the better both
algorithms are compared to CGA.
For a numerical comparison of CGA and MRA in this respect see
\cite{desy92111}.
The numerical results discussed below refer to a $4^3 \cdot 8$ lattice
and an error bound of $\delta = 10^{-8}$.
\end{enumerate}

\subsection{Direct parameters}
Now let us consider the direct parameters.
{}From the prescription for the odd-even preconditioning it follows that
it works the better the smaller $K$ and the larger the Yukawa couplings
are.
For such values of the direct parameters MRA will always be the best
choice.
It will be better than CGA, because preconditioning works well, and
better than BGA, because it needs roughly half the amount of
arithmetic for one iteration compared to the latter.
If, on the other hand, the Yukawa couplings are small and the fermionic
hopping parameter is large, an improvement by this kind of
preconditioning is not possible.
Therefore CGA will be faster than BGA and MRA in this case.
The most favourable parameter region for BGA is thus expected to be at
intermediate values of $K$, where MRA begins to break down and CGA slows
down.
Furthermore, when the Yukawa couplings strongly differ from each other,
i.e.\ one of them has a large absolute value, while the other Yukawa
coupling has a small one, then this is unfavourable for both CGA and MRA.

In order to check the expectations stated above I have made several
tests.
If the scalar field is uniform CGA is the best algorithm for all
parameter values under investigation.
The other extreme consists of a randomly chosen scalar field.
In this case it can be seen from figure 1 that for suitably chosen
values of $G_\psi$ and $G_\chi$ BGA is by far the fastest algorithm in a
broad region of intermediate $K$.
For disadvantageous values of the Yukawa couplings this $K$-region can
shrink to zero and we have a situation like that depicted in figure 2,
where for each value of $K$ CGA is better than BGA.
An intermediate case is shown in figure 3.
Here for small values of $K$ BGA is faster than CGA, but it is exceeded
by MRA.
It follows a small region with nearly the same performance of BGA and
CGA, before at larger values of $K$ CGA again becomes faster.
This figure comes from a calculation within the SU(2)-version of the
model.
Of course the extreme cases, depicted for the U(1)-version
in figure 1 and 2, are also existent in the SU(2) model.
{}From the general discussion above we expect that the relative behaviour
of the algorithms is the same for the SU(2) model and for the U(1)
model.
For this reason and in order to save computer time the following tests
are done within the U(1)-version.

The dependence on the Yukawa couplings can be read off from figure 4.
There all parameters except $G_\chi$ are fixed.
When $G_\chi$ moves away from $G_\psi$ to larger values, only BGA
remains fast, whereas CGA slows down considerably and for MRA convergence is
not observed beyond a certain value of $G_\chi$-value.

\subsection{Indirect parameters}
Considering the indirect parameters, $\kappa$ is the most interesting one,
because no large effects from $\lambda$ are expected even if it is
varied over a large range.
If all other parameters are held fixed it depends on $\kappa$ whether
the system is in the PM, FM, AFM or FI phase \cite{desy}, i.e.\ it
determines the magnetisation $|\phi|= (1/N) |\sum_x \phi_x|$, where the
sum extends over the $N$ lattice sites.
For BGA the amount of time for one inversion is of the same magnitude,
no matter whether we set $\phi$ randomly ($|\phi| \approx 0$) or choose
a uniform configuration ($|\phi| = 1$).
The performance of CGA, however, is strongly influenced by the
magnetisation.
CGA needs, depending on the other parameters, up to 60 times longer for
a random configuration, than for a uniform one.
(This is the largest factor found during the tests, but probably not the
largest possible.)
Concerning this aspect MRA shows opposite behaviour to CGA.
It needs more time for inverting a matrix with uniform configuration.

These results suggest that BGA is the most promising candidate in
regions of intermediate magnetisation.
In fact, this expectation is confirmed by table 1.
The table shows the time requirements for one inversion, needed by the
different algorithms.
Three types of scalar field configurations are considered: random,
uniform and equilibrated by 500 trajectories for
$\kappa= 0.0$ and $\kappa = \pm 0.1$.
This values of $\kappa$ cover the FM, PM and AFM phase \cite{desy}.

The relative speed of algorithms in table 1 depends only little on the actual
scalar field configurations. One can observe, following the development
of magnetisation and time requirements for the inversion during the
equilibration, that at some value of $|\phi|$, BGA will become faster
than CGA for suitable $\kappa$ (see also table 3 for another set of
parameters).
For the particular parameters of table 1 BGA represents the best
inversion method only for a small region of $|\phi|$, i.e.\
$\kappa$.
Moreover, these values of $\kappa$ are of little physical interest,
because the system is not near a phase transition.

The picture changes drastically if we choose $G_\psi$ to be zero,
instead of $0.1$.
This choice is physically important since it represents the decoupling
scenario \cite{desy92111,Desy 91}, which requires $K=0.125$ in addition.
The time requirement for one inversion then increases by a large
factor for all three algorithms, but for suitable $K$ this results in
an advantage for BGA.
It is the fastest algorithm for all physically interesting
magnetisations, and is beaten by CGA only deeper in the FM phase.
To demonstrate this I consider $K=0.125$ and three values of $\kappa$.
Two of them are placed near the second order phase transition between
the FM and PM phases ($\kappa=-0.04,-0.06$), the third one at the
transition between the PM and AFM phases ($\kappa=-0.173$).
The results on the algorithmic speeds are summarized in table 2.
The development of the time requirements and of the magnetisation
during the equilibration are shown in table 3.
One observes that after some equilibration BGA is the best choice for
this set of parameters, if the magnetisation is not too large.
The MRA did not show convergence or was very slow for all values of $\kappa$.

The discussion above shows that three different cases occur:
\begin{enumerate}
\item If the values of $G_\psi,G_\chi,K$ and $\kappa$ are most
favourable for BGA, then after equilibration, starting from an uniform
configuration, BGA is the fastest algorithm and has to be used for the
production runs.
\item If the direct parameters have somewhat less suitable values, there
are still values of $\kappa$ for which BGA is faster than CGA, but if
the magnetisation is too small, it will be beaten by MRA (see table 1).
In such a situation it might happen that BGA is the best method of
inversion only in a small, physically uninteresting region or even for
no value of $\kappa$ at all.
\item In the third case the direct parameters are not appropriate for
odd-even preconditioning and CGA will always be the fastest algorithm,
no matter how large $\kappa$ is.
\end{enumerate}

\section{Summary and Conclusion}
The biconjugate gradient algorithm (BGA) for the inversion of large
matrices, which combines advantages of the conjugate gradient (CGA) and
minimal residue algorithms (MRA), has been investigated.
In the framework of Higgs-Yukawa models it turns out that there exist
regions in parameter space where the other two algorithms slow down and
BGA represents an improvement.
To be precise, BGA can be the optimal choice in this model, if the
fermionic hopping parameter assumes intermediate values, such that the
fermionic mass is near zero, and if both Yukawa couplings differ
strongly in magnitude, as is the case in the decoupling situation.
Furthermore the magnetisation must not be too large, because otherwise
CGA is the best choice.
For small magnetisations it depends on the particular choice of direct
parameters, whether BGA is beaten by MRA.

In those parameter regions where BGA is the fastest algorithm,
using higher order preconditioning for BGA does not result in a further
gain.
Similar to the case of MRA, going to larger lattices or smaller error
bounds makes BGA more attractive.

Two further things should be mentioned here.
For some Monte Carlo algorithms, like e.g.\ the Langevin algorithm,
the updates require the inversion of $Q$ only, instead of $Q^+Q$.
This reduces the amount of time for BGA or MRA inversions by a factor
of two, which is of course not the case for CGA.

Secondly, taking the matrix $Q$ or $Q^+$ itself, without
preconditioning, i.e.\ without odd-even decomposition, as input for
BGA or MRA no gain is found compared to CGA, in contrast to the
suggestion of \cite{rec}.
Maybe this is due to the special form of our matrix.\\
\\

{\bf Acknowledgement}\\

\noindent
I would like to thank Prof.~G.~M\"unster and Dr.~L.~Lin for careful reading
of the manuscript and useful comments.\\
All numerical calculations have been performed on the CRAY Y-MP of HLRZ,
J\"ulich.
%

%%%%%%%%%%%%%%%%%%%%%%%%%%%%%%%%%%%%%%%%%%%%%%%%%%%%%%%%%%%%%%%%%%%%%%%
\newpage
%%%%%%%%%%%%%%%%%%%%%%%%%%%%%%%%%%%%%%%%%%%%%%%%%%%%%%%%%%%%%%%%%%%%%%%
\section*{Figure captions}

\noindent
{\bf Fig. 1.} Comparison of BGA, CGA and MRA for the U(1)-model on a
$4^3 \cdot 8$ lattice with random scalar field configuration.
The parameters are $\lambda=10, \kappa=0, G_\psi=0.1, G_\chi=1.0$.
The figure shows the time needed for a matrix inversion in seconds
versus the fermionic hopping parameter.
The symbols are crosses (BGA), squares (MRA) and diamonds (CGA).\\

\noindent
{\bf Fig. 2.} The same as figure 1, but at $G_\chi=0.3$\\

\noindent
{\bf Fig. 3.} Comparison of matrix inversion algorithms for the
SU(2)-model at $\lambda=10^{-6}, \kappa=0.0, G_\psi=0.3, G_\chi=0.0$ on
a $4^3 \cdot 8$ lattice with a random scalar field configuration.
The same symbols as in figure 1 are used.\\

\noindent
{\bf Fig. 4.} Comparison of matrix inversion algorithms at fixed $K$ and
varying $G_\chi$ for the U(1)-model on a $4^3 \cdot 8$ lattice with a
random scalar field configuration.
The parameters are $\lambda=10, \kappa=0, G_\psi=0.0, K=0.125$.
The figure shows the time needed for a matrix inversion in seconds
versus the Yukawa coupling $G_\chi$.
The symbols are as in figure 1.
%%%%%%%%%%%%%%%%%%%%%%%%%%%%%%%%%%%%%%%%%%%%%%%%%%%%%%%%%%%%%%%%%%%%%%%
\newpage
%%%%%%%%%%%%%%%%%%%%%%%%%%%%%%%%%%%%%%%%%%%%%%%%%%%%%%%%%%%%%%%%%%%%%%%%
\begin{table}
\caption{Time in seconds for one inversion of the fermion matrix at
         different magnetisations (random, equilibrated at
         $\kappa=-0.1,0.0$ and $+0.1$, and uniform).
         The parameters are $G_\psi =0.1, G_\chi=-1.0, \lambda=1,
         K=0.125$.
         The numbers for $|\phi| \approx 0$ are averages
         over random configurations.}
\begin{center}
%{\bf time (s) for one inversion at different magnetisations}\\
\begin{tabular}{|c||r|r|r|}
\hline
$|\phi|$   &  CGA   & BGA  & MRA  \\
\hline
$\approx 0$  &  0.882 & 0.255 &  0.123 \\
0.084        &  0.67  & 0.42  &  0.267 \\
0.275        &  0.61  & 0.45  &  1.05  \\
0.83         &  0.368 & 0.443 &  3.11  \\
1            &  0.23  & 0.34  &  3.67  \\
\hline
\end{tabular}
\end{center}
\end{table}
%%%%%%%%%%%%%%%%%%%%%%%%%%%%%%%%%%%%%%%%%%%%%%%%%%%%%%%%%%%%%%%%%%%%%%%%
\begin{table}
\caption{Time in seconds for one inversion of the fermion matrix at
         different magnetisations (random, equilibriated at
         $\kappa=-0.173, -0.06$ and $-0.04$, and uniform).
         The parameters are $G_\psi =0.0, G_\chi=-1.0,\lambda=10$.
         Empty places mean that the MRA did not show convergence within
         an acceptable amount of iterations.
         The numbers for $|\phi| \approx 0$ are averages
         over random configurations.}
\begin{center}
%{\bf time (s) for one inversion at different magnetisations}\\
\begin{tabular}{|c||r|r|c|}
\hline
$|\phi|$  & CGA   & BGA  & MRA  \\
\hline
$\approx 0$ & 13.720 & 1.650 & ---   \\
0.07        &  1.040 & 0.630 & 1.49  \\
0.168       &  0.800 & 0.640 & 4.50  \\
0.259       &  0.547 & 0.622 & ---   \\
1           &  0.282 & 0.379 & ---   \\
\hline
\end{tabular}
\end{center}
\end{table}
%%%%%%%%%%%%%%%%%%%%%%%%%%%%%%%%%%%%%%%%%%%%%%%%%%%%%%%%%%%%%%%%%%%%%%%%
\begin{table}
\caption{ Time in seconds for one Hybrid Monte Carlo trajectory,
          averaged over the first $n=100$ trajectories starting with a
          uniform configuration, over the next $n=300$ trajectories, and
          over the last $n=300$ trajectories.
          The parameters are the same as for table 2 and the step size
          is $\epsilon = 0.04$.
          $<|\phi|>$ is the magnetisation averaged over $n$
          measurements.}
\begin{center}
%{\bf time (s) for n trajectories starting with uniform configuration}\\
\begin{tabular}{|c|c||r|r|r|}
\hline
$\kappa$ & $<|\phi|>$ & $n$  & CGA   & BGA  \\
\hline
         & 0.51       & 100      & 176   & 434  \\
-0.04    & 0.23       & 300      & 1270  & 1470 \\
         & 0.23       & 300      & 1230  & 1452 \\
\hline
         & 0.49       & 100      & 174   & 413  \\
-0.06    & 0.205      & 300      & 1340  & 1450 \\
         & 0.175      & 300      & 1510  & 1480 \\
\hline
         & 0.37       & 100      & 229   & 407  \\
-0.173   & 0.085      & 300      & 2120  & 1490 \\
         & 0.068      & 300      & 2320  & 1560 \\
\hline
\end{tabular}
\end{center}
\end{table}
%%%%%%%%%%%%%%%%%%%%%%%%%%%%%%%%%%%%%%%%%%%%%%%%%%%%%%%%%%%%%%%%%%%%%%%%%%%
\end{document}